# "Smoking gun" signatures of topological milestones in trivial materials by measurement fine-tuning and data postselection

S.M. Frolov[1&], P. Zhang[1*%], B. Zhang[1*], Y. Jiang[1*], S. Byard[1*], S.R. Mudi[1], J. Chen[2]

*1 Department of Physics and Astronomy, University of Pittsburgh, Pittsburgh, PA 15260, USA*

*2 Department of Electrical and Computer Engineering, University of Pittsburgh, Pittsburgh, PA 15260, USA*

A.-H. Chen[3], M. Hocevar[3]

*3 Univ. Grenoble Alpes, Grenoble INP, CNRS, Institut Néel, 38000 Grenoble, France*

M. Gupta[4], C. Riggert[4], V.S. Pribiag[4]

*4 School of Physics and Astronomy, University of Minnesota Twin Cities, Minneapolis, Minnesota 55455, USA*

\* These authors contributed equally to this work.

& Corresponding author frolovsm@pitt.edu
% Current address: Beijing Academy of Quantum Information Sciences, 100193 Beijing, China

September 18, 2023

**Abstract**

Exploring the topology of electronic bands is a way to realize new states of matter with possible implications for information technology. Some states, such as several types of topological insulators, and quantum Hall phases, have been established. Others, such as topological and triplet superconductors remain a focus of an engaging search which includes a diversity of approaches. Because bands cannot always be observed directly, a central question is how to tell that a topological regime has been achieved. Experiments are often guided by a prediction of a unique signal or a pattern, called "the smoking gun". When a "smoking gun" is observed, this is taken as unambiguous proof of a new state of matter. Examples include peaks in conductivity, microwave resonances, and shifts in interference fringes. However, many condensed matter experiments are performed on relatively small, micron or nanometer-scale, specimens. These structures are in the so-called mesoscopic regime, between atomic and macroscopic physics, where phenomenology is particularly rich. In this paper, we demonstrate that the trivial effects of quantum confinement, quantum interference and charge dynamics in nanostructures can reproduce accepted smoking gun signatures of triplet supercurrents, Majorana modes, topological Josephson junctions and fractionalized particles. The examples we use correspond to milestones of topological quantum computing: qubit spectroscopy, fusion and braiding. None of the samples we use are in the topological regime. The smoking gun patterns are achieved by fine-tuning during data acquisition and by subsequent data selection to pick non-representative examples out of a fluid multitude of similar patterns that do not generally fit the "smoking gun" designation. Building on this insight, we discuss ways that experimentalists can rigorously delineate between topological and non-topological effects, and the effects of fine-tuning by deeper analysis of larger volumes of data.

# Introduction

Electrons in solids are restricted in their energies to intervals called bands. By changing the composition or structure of crystals it is possible to re-order the bands and realize a non-trivial band topology. Topological bandstructures are often accompanied by edge states(*1*). In two-dimensional materials, they can be chiral or helical propagating modes. In one-dimensional wires, they are stationary quantum confined end states. The edge states may have interesting properties, such as the predicted non-Abelian exchange rules(*2*). Proposed applications in electronic and quantum technology, such as dissipationless transfer and robust storage and processing of information(*3*), drive a lot of the interest in topological materials.

Many significant discoveries in physics are made through observing a signal, or a pattern, that is so unique that it eliminates or dramatically reduces alternative interpretations. Examples from solid state physics include plateaus of conductance that established the quantum Hall effect(*4*), half-period shifts in superconducting interference patterns (*5*) which demonstrated the d-wave symmetry of the superconducting wavefunction in cuprates, anticrossing in the microwave spectrum demonstrating vacuum Rabi splitting and strong coupling regime in microwave resonators(*6*). It is not without reason that evidence for new topological states is sought in the form of smoking gun signatures. For instance, linearly dispersing bands forming a so-called Dirac cone are directly visible in angle-resolved photoemission measurements on 3D topological insulator crystals(*7*).

At the same time, the smoking gun-centered approach poses certain challenges. Often the identification of a measurement as a smoking gun is based on a qualitative argument. This leaves room for bias, especially when the signal is less dramatic, or the signal-to-noise ratio is low. Predicted smoking guns are frequently drawn from simplified models that do not include all the factors present in real materials and experiments. Unusual and unexplained signals are sometimes treated as smoking guns with an explanation put together post-hoc. Control knobs are fine-tuned, sometimes narrowly, until a desired pattern is observed. This creates a risk of zeroing in on chance fluctuations that imitate expected data when no underlying physics is present. Finally, data presented in the publication are necessarily narrowly selected from a larger set, introducing the chance of selection bias.

In this paper, we focus on how fine-tuning during data acquisition and subsequent data selection in mesoscopic samples can generate believable smoking gun signatures corresponding to milestones in topological quantum computing(*8*). A topological qubit operates on particles called anyons, to distinguish them from fermions and bosons. While no topological qubits have been demonstrated, the new breakthrough physics required to attain them includes the spectroscopic evidence of anyons, their fusion to read out the state of the qubit as well as braiding to perform logic gates. Anyons may be found in topological superconductors or in fractional quantum Hall systems.

First, in nanowire junctions we demonstrate Josephson supercurrent increasing with increasing magnetic field, suggestive of triplet Cooper pairing, a phenomenon often linked to topological superconductivity(*9*). The behavior is observed in a particular regime and is qualitatively explained by finite-bias resonances in the junction suppressing switching current around zero field.

Second, as a smoking gun of Majorana end states in nanowires(*10, 11*), which are predicted to be anyons, we demonstrate plateaus in conductivity at zero bias in tunneling devices(*12*). The zero-bias plateaus are obtained at zero external magnetic field which would be a major advantage for topological quantum computing. However, the signal is a result of fine-tuning of non-topological Andreev bound states in a quantum dot with the assistance of remote micromagnets.

Third, in superconductor-semiconductor quantum well Josephson junctions, we demonstrate a hallmark of the fractional Josephson effect, which is associated with the fusion of Majorana zero modes. We observe the disappearance of several odd-order Shapiro steps under microwave irradiation(*13*). The effect is obtained in the trivial regime where no Majorana modes are expected and is a result of fine-tuning.

And finally, in semiconductor nanowire quantum dots, we find examples of Coulomb blockade patterns shifted by fractions of a period, including by approximately 1/3. A tempting interpretation of these data would be that they result from adding fractional charges to the island(*14*). Fractional shifts can serve as proof of anyons and in the anti-dot configuration they can in principle serve as demonstration of anyon braiding. However, such charges are only expected at large magnetic fields, in the fractional quantum Hall effect, whereas our quantum dots are studied at zero applied field. A less exotic explanation is that these shifts are induced by moving integer charges in a nearby charge trap that is capacitively coupled to the quantum dot.

We conclude by summarizing the limitations of the smoking gun approach and propose ways to enhance verifiability and robustness of research claims.

**Example 1 – "Triplet supercurrent" in Sn/InSb nanowire junctions**

Superconductivity has fascinated researchers for over a century. Discoveries on this topic come with impressive regularity. We have also seen situations where unusual observations are ascribed to superconductivity but turn out unrelated(*15*, *16*), or when unexplained features in superconducting samples are believed to be exotic types of superconductivity(*17*, *18*), albeit without a confirmation(*19*, *20*).

Superconductivity is based on the coupling of electrons known as Cooper pairing. When it comes to the spin of a pair, the only two options are singlet and triplet. Most, if not all, known superconductors are singlet superconductors. There is intense research into whether triplet superconductors can be created and verified(*21–23*). Topological superconductors which host anyon states are one type of triplet superconductors(*24*). One basic expectation for a triplet superconductor is that it could get enhanced in an external magnetic field. Indeed, the magnetic field promotes the alignment of spins, and favors triplet over singlet.

We study superconductor-semiconductor nanowire Josephson junctions, based on InSb nanowires with Sn shells(*25–27*), controlled by electrostatic gates. These structures are believed to be candidates for a type of triplet topological superconductivity induced in the semiconductor nanowire by proximity in the presence of applied magnetic field(*28*). As evidence of induced superconductivity, we observe dissipationless Josephson supercurrent(*26*). In contrast with most reports to date, here supercurrent is found to be increasing when magnetic fields are applied, nearly parallel to the nanowire. In Fig. 1A, supercurrent starts at a value of 0.5 nA near zero field. However, at substantial fields of approximately 0.6T the signal reaches over 1nA, almost double the zero-field value. The level of the signal is well above the noise floor of our measurement, which is of the order of 10 pA for detecting a switching current. For a standard singlet superconductor, we expect such substantial magnetic fields to gradually suppress superconductivity. This eventually happens at around B=2.0T where the Sn layer itself stops superconducting thereby suppressing the induced superconductivity as well (Fig. 1B). Fig. 1C demonstrates that the effect persists over a finite range of gate voltages which tune the Josephson junction, and thus represents a reproducible and robust observation. Measurements of superconductivity and supercurrents that are enhanced by magnetic field were interpreted as evidence of exotic superconductivity in previous works(*9*, *22*).

At the same time, the gate voltage range over which the field-enhanced supercurrent is found is limited (Fig. 1C). Outside this range, for more negative and more positive gate voltages, supercurrents exhibit an initial decay with magnetic field, sometimes followed by oscillations due to quantum interference. While we found examples of supercurrent increasing above the zero-field value in a few devices, most samples exhibit a maximum supercurrent at zero field. In the same nanowire junction, in the single-subband regime where transport of electrons is fully spin-polarized, we found no supercurrents at high magnetic field, consistent with the absence of triplet pairing(*26*). This pushes us to search for explanations that do not involve triplet or other types of exotic superconductivity.

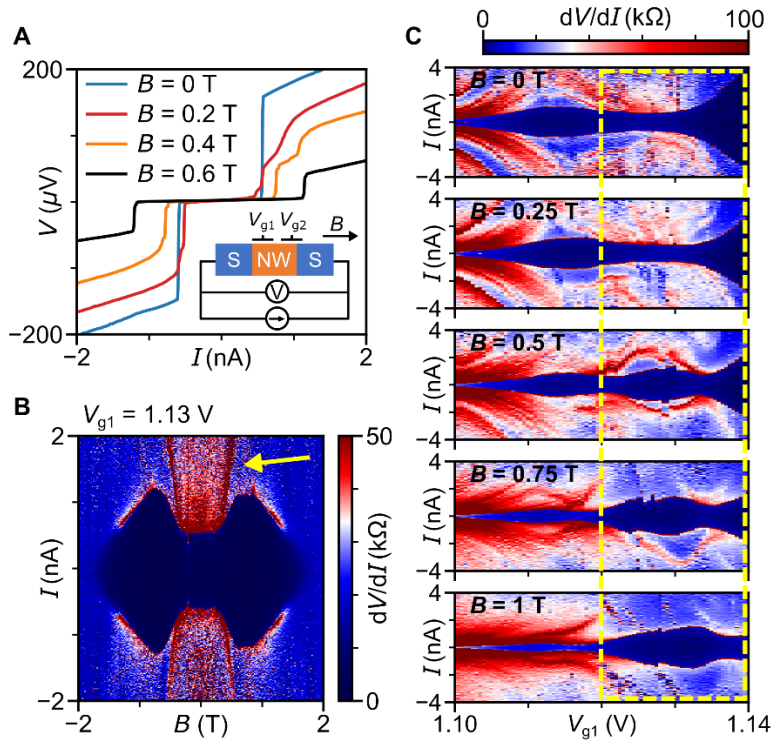

**Figure 1. Magnetic field enhanced supercurrents in InSb-Sn nanowire junctions.** A) Current voltage I-V characteristics at different magnetic fields B. Inset shows device measurement diagram. B) Differential resistance in magnetic field and current bias. Arrow points at a finite voltage resonance. C) Gate voltage evolution of differential resistance for several magnetic fields, marked in the upper left corner of each plot. Yellow dashed lines denote region where supercurrent is enhanced with B. All measurements at T = 50 mK, B is parallel to the nanowire. The second gate $V_{g2}$ = -2.5V. All the scans are done by sweeping from zero bias to the positive and negative bias.

We notice that highly non-monotonic supercurrents are frequently accompanied by sharp peaks in the finite voltage state, located outside the supercurrent window (Fig. 1B). They originate from mesoscopic features of the nanoscale junctions, such as multiple Andreev reflections, Andreev bound states, self-induced Shapiro steps. We observe that the dependence of the switching current on the magnetic field is dependent on the behavior of mesoscopic resonances. If one of these peaks moves to higher voltage in magnetic field, it leads to a larger switching current. The explanation does not involve the unusual triplet pairing and suggests that supercurrents growing with field do not prove exotic superconductivity. Instead, changes in differential resistance cause switching in and out of the finite voltage state by altering the phase dynamics in the junction. Beyond the mesoscopic explanation given here, magnetic field enhanced supercurrents in nanowires were attributed to the polarization of magnetic impurities and to quasiparticle dynamics(*29–31*).

**Example 2 – "Majorana" zero bias peak plateau**

There are other ways to demonstrate the type of superconductivity which we are interested in, the topological kind, through the observation of Majorana bound states. In nanowire devices, Majorana states are most frequently investigated in tunneling experiments which yield conductance peaks at zero bias that appear at finite applied magnetic field(*32*). In the language of topological qubits, these measurements are akin to qubit spectroscopy. It is also known that topological states are not the only effect that generates zero-bias peaks. In fact, mesoscopic physics knows many different zero-bias conductance anomalies, from Kondo effect to Andreev bound states(*33, 34*).

In order to connect the observation of zero-bias peaks with Majorana physics, it is important to look at additional features of the signal. If the data exhibit those it can be perceived as the smoking gun. For instance, a zero-bias conductance plateau has been predicted for zero-bias tunneling peaks due to Majorana(*12*). The basic idea is that peaks due to other origins are transient while Majorana results in fixed conductance over an extended parameter range, ideally of a quantized value in the units of $2e^2/h$.

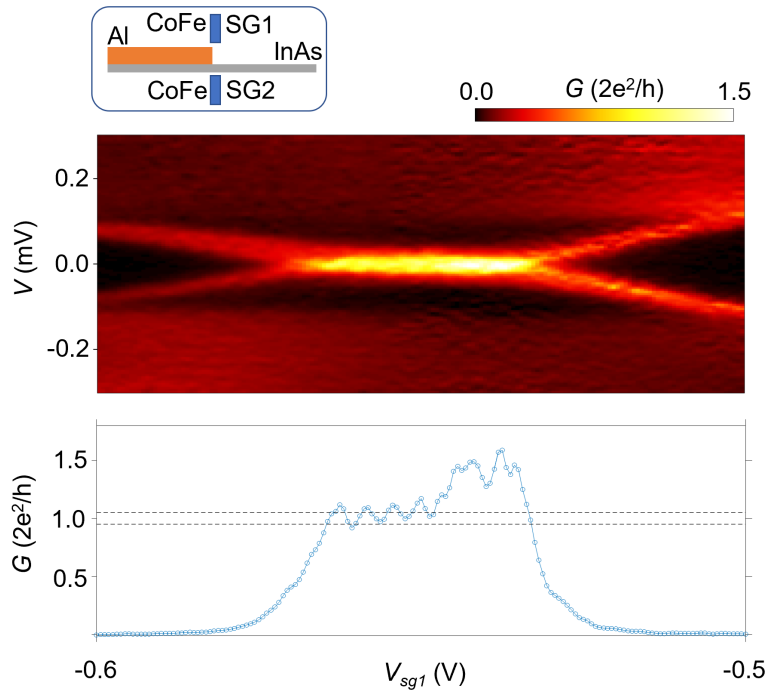

**Figure 2. Zero-bias conductance plateau in InAs-Al nanowire tunneling device.** A) A scan of bias voltage V vs. side gate voltage $V_{sg1}$. The back-gate voltage is 6.075 V and the second side gate voltage $V_{sg2}$ is 0 V. Diagram on top is a device schematic, device image is found in supplementary materials. The CoFe side gates are magnetized first and then the external field is set to zero. B) Line-cut at zero voltage bias. A series resistance of 15795Ω was subtracted to get the zero-bias conductance close to the quantized value. Dashed lines indicate the +/- 5% range of conductance around the quantized value.

We search for zero-bias conductance plateaus(*35*) in InAs-Al nanowire devices fabricated with one non-superconducting contact, in the tunneling geometry. We identify regimes where, over a finite range of gate voltages, conductance at zero voltage bias changes weakly (Fig. 2). What is also interesting about these data is that they are obtained at zero external magnetic field(*36*). However, on-chip magnetic fields are present due to micromagnets positioned near the tunnel junctions. This geometry offers advantages for Majorana experiments allowing for varied field profiles on the nanoscale, not available with large solenoid coils(*37*).

We also demonstrate conductance near the value of $2e^2/h$ on part of the plateau. However, in other regimes, we observe similar plateaus with values exceeding $2e^2/h$, and this contradicts the Majorana explanation. The plateau-shaped appearance of the signal is a result of fine-tuning to an expected pattern. In other regimes, peaks that look overall the same do not appear as plateaus. Furthermore, the value of conductance reported in a figure is sensitive to a subtracted series resistance, which can be chosen to bring the value closer to the quantized value(*38*).

Despite their superficially promising appearance, these "plateaus" cannot be connected to Majorana modes for a number of reasons. First, the fact that a pattern that is visually strikingly similar is reported at $2e^2/h$(*39*), and at a lower or higher values, voids the interpretation in terms of theoretically predicted quantized Majorana conductance. Second, there exists a more likely explanation for these effects being due to Andreev bound states in unintended quantum dots near the tunnel barrier. The dots are quantum confined regions created due to complex geometry and disorder. Given that Andreev bound states are expected in such devices, it is important to try to exclude them as the origin of the observed signal prior to suggesting a Majorana origin. We present evidence of quantum dots in our devices in supplementary information, similar data are likely to be found in most devices that exhibit zero-bias peaks experimentally. The open dots, with very low tunneling barriers are difficult to identify through Coulomb blockade, but they can still host localized states strongly coupled to nearby superconductors. Given a well-developed theoretical foundation, it should be possible to identify Majorana zero modes based on zero-bias peaks measurements, through critical and comprehensive analysis of larger data sets. Though a pre-declared set of criteria for what constitutes a Majorana observation can also be emulated with trivial mesoscopic effects(*40, 41*).

**Example 3 – "Fractional Josephson effect"**

Essential for a Majorana-based topological qubit is the fusion process which occurs when two Majorana states are brought together(*8*). Being their own antiparticles, they annihilate. The zero-energy modes move to finite energy and reveal their occupation: either an electron or a hole, the outcome corresponds to the readout of the qubit. In order to test whether Majorana fusion is taking place, we return to measurements on Josephson junctions. If the two superconductors of the junction are topological, this would result in two Majorana modes located across the junction barrier. In order to bring the two Majorana modes together and pull them apart, we wind the superconducting phase difference across the junction. At the phase difference of exactly π the two modes are decoupled while at any other phase difference they are partially or completely fused(*42*). This is known as the fractional Josephson effect because it appears that supercurrent is mediated by fractions of Cooper pairs, their halves, and the Josephson relation is proportional to $\sin(\varphi/2)$.

The most experimentally accessible way to search for the fractional Josephson effect is by studying the staircases of the so-called Shapiro step resonances(*13, 43–45*). The phenomenon is common to all Josephson junctions where under external radio frequency (rf) irradiation the current-voltage characteristics develop distinctive steps at voltages in multiples of h*f*/2e, where *f* is the rf frequency. In the case of the fractional Josephson effect, every second Shapiro step should disappear, with steps only at multiples of h*f*/e (Fig. 3A).

In planar Josephson junctions fabricated out of InAs quantum wells coated with Al superconductor(*46*), we observe a characteristic pattern of missing Shapiro steps, where steps at h*f*/2e, 3h*f*/2e and 5h*f*/2e are not present in current-voltage characteristics, while steps at h*f*/e, 2h*f*/e and 3h*f*/e are present (Figs. 3B, 3C)(*47*). The missing steps are observed over a significant range of frequencies of several gigahertz, though steps gradually re-appear at higher frequencies. Given these additional indicators of the robustness of the effect, the observation is consistent with a smoking gun fractional Josephson effect(*13*).

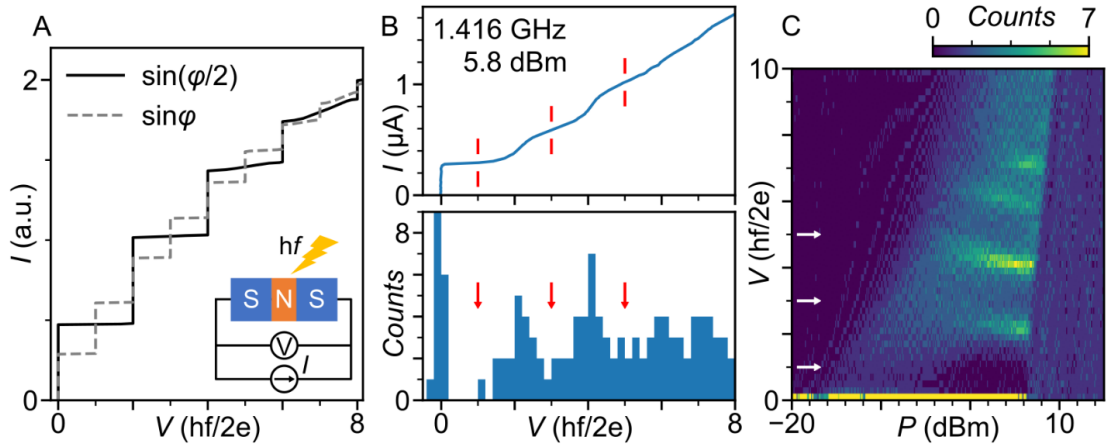

**Figure 3. Missing odd-order Shapiro steps in planar Josephson junctions.** A) Expected Shapiro step spectrum for regular (dashed line) and fractional (solid line) Josephson effects. Inset shows measurement diagram. B) Current voltage characteristic and voltage histogram with missing Shapiro steps indicated by red lines and arrows C) Power-voltage histogram with missing steps indicated by white arrows.

However, the junctions are measured near zero magnetic field, and therefore are not in the topological regime. No Majorana modes are expected in these devices under these conditions. Moreover, in other regimes we observe missing steps at some of the even frequencies, as well as additional steps at half-frequencies. We also note that wide planar junctions are not the right system to observe the fusion of just the two Majorana modes, because they are populated with hundreds of modes lined up along the junction. If a single mode at the edge becomes Majorana, that would constitute a small percentage contribution to the overall Josephson effect and not lead to the disappearance of Shapiro steps.

By way of explanation, we notice that the missing steps are observed at relatively low frequencies, where rounding of the Shapiro steps makes them less distinct. We argue that other non-linearities in current-voltage characteristics can disrupt the regular step patterns, and with enough fine tuning, lead to the visual disappearance of only the odd steps. A model we developed illustrates how this can be possible assuming bias-dependent resistivity. Other works found missing first-order Shapiro steps(*43*, *48*) in non-topological regimes and explained this in terms of heating and Landau-Zener transitions(*49*, *50*). Overall, the mere observation of an even Shapiro step pattern does not appear to serve as the smoking gun evidence of the fractional Josephson effect.

**Example 4 – "Fractional charge" jumps in nanowire quantum dots**

Anyon objects or quasiparticles come with fractionalized properties. For instance, two Majorana modes can be thought of as an electron split between two ends of a nanowire. Anyons known from the quantum Hall effect are characterized by fractional charges, such as 1/3 of the elementary charge(*51*). In principle, conductance plateaus at values such as 1/3 or 4/3 in the fractional quantum Hall regime already provide strong evidence of fractional charge quasiparticles(*52*). However, a variety of experiments search for direct evidence of fractional charges through noise spectroscopy, interferometry and charge sensing(*53–55*). These experiments, especially when done in the interferometer configuration and in the even denominator fractional quantum Hall regime(*56*), are useful for topological quantum computing because they represent anyon braiding, the logic gate operation on the topological qubit.

One way to search for fractional charges is by analyzing periodic patterns of conductance. In interferometers or isolated charge islands, conductance can exhibit oscillations periodic in charge or magnetic flux or both. In Figure 4A, we demonstrate an example of Coulomb blockade patterns in a semiconductor quantum dot, which are stripes of high conductance followed by stripes of low conductance, controlled by two electrostatic gate voltages. Moving from one stripe to another is equivalent to adding a single charge to the island.

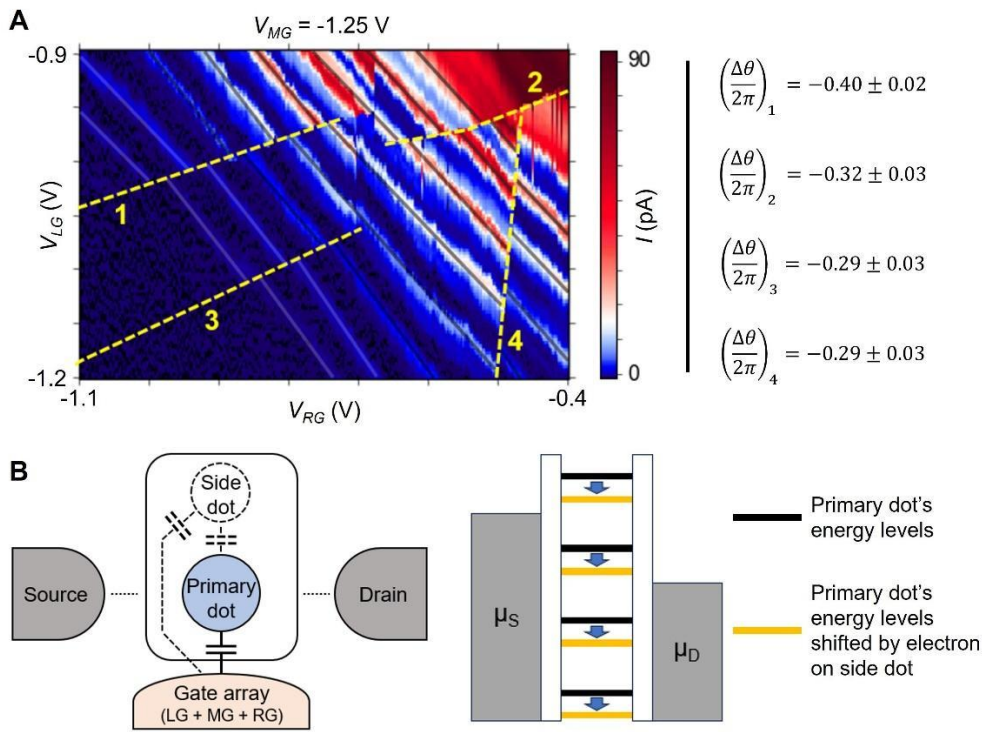

**Figure 4. Coulomb resonance spectroscopy in a nanowire quantum dot.** A) Gate voltage dependence of current. Dashed lines indicate pattern shifts, solid lines along the conductance stripes are guides to the eye. Period shifts calculated for each line are indicated to the right of the panel with subindices corresponding to the dashed lines. B) Model for a primary dot capacitively coupled to a side dot, a shift in the side dot charge shifts energy levels in the primary dot.

In addition to the regular diagonal stripe pattern, we observe abrupt shifts upon which the pattern skips by a fraction of the period. Many of the jumps correspond to approximately shifts of 1/3 of the period. This is characterized by a value Δθ/2π, where a shift by a full period would be equal to 1. The shifts are not vertical or horizontal, which could be explained by noise on either of the gate electrodes, but depend on both the gates, indicating that they are the property of the device under investigation.

One explanation for these jumps is that charges of non-integer value are added stochastically(*53*) to the quantum dot. However, there is no reason to expect fractional charges in this regime, because the quantum dot is measured at zero applied magnetic field, away from the fractional quantum Hall regime. Furthermore, it is created in a quasi-one-dimensional semiconductor PbTe nanowire(*57*) while quantum Hall effects are observed in two-dimensional layers. Instead, the explanation of these jumps is that near the quantum dot under investigation there is another, capacitively coupled dot (Fig. 4B). The side dot is connected in parallel. Current either does not pass through this dot or the current through it is much lower than through the main dot. However, mutual capacitance between the dots is such that an addition of one electron to the side dots induces approximately 1/3 of the charge on the primary dot. When this is done controllably, by intentionally fabricating an additional coupled quantum dot, this is in fact one of the common approaches for sensing the charge occupation in electron-spin qubits(*58*).

However, uncontrolled dots or charge traps are frequent in nanoscale devices where metallic surfaces and dielectric layers are in close proximity to current paths. Gate leakage can provide another pathway for charges entering quantum dots(*59*). The circumstances described above are also present in interferometer devices which are studied in the quantum Hall regime. There, localized areas within the interferometer are a few hundred nanometers in size and are surrounded by gate electrodes, dielectric layers and by adjacent conducting areas. Charges moving in nearby charge traps can affect the interference patterns. Evidence of anyon braiding should extend beyond the demonstration of charge jumps to other signatures, such as through noise correlations, though care should be taken to ensure those cannot be emulated with the help of mesoscopic effects.

**Discussion and Conclusion**

Through these four experiments, we demonstrate that mesoscopic physics is rich enough to provide a remarkable variety of striking patterns. Many of them may fit with the expectations for unusual phenomena, but at the same time have a different origin. Beyond these examples, a reader is encouraged to check out other recent works dedicated to alternative interpretation of smoking gun signatures. In nanowires with a full superconducting shell, a coincidence of zero-bias conductance peaks and Little-Parks oscillations was demonstrated to be accidental and not due to Majorana modes induced by odd vorticity(*60, 61*). In the same devices, researchers have shown that earlier claims of Coulomb peak spacing oscillations having a relation to Majorana physics are not justified(*62, 63*), because oscillations are strongly dependent on fine-tuning of irrelevant parameters. Conductance quantization can be adjusted by subtracting series resistance, and this was shown to be used to produce perfectly quantized values of conductance(*38*).

What unites these examples is the impossibility of identifying new states of matter from a single graph, using a "smoking gun" argument, when studying samples with strong potential for mesoscopic physics such as quantum confinement and interference, defects and disorder, complex geometry, and uncontrolled dynamics of carriers. It is important to vary the experimental parameters over a wider range to see whether the behavior persists, and publishing such larger datasets can inoculate against the possibility that smoking guns will turn out to be artefacts of fine-tuning.

Crosschecks and secondary signatures, that detail the behavior of smoking gun signals, are often added such that the search is focused on a recognizable pattern rather than a single value of a signal. Though even those can be sometimes mimicked through fine-tuning. In some of the cases we discuss here, the similarity between expected topological behavior and data is superficial and can be easily spotted with a small amount of additional data. In other cases, such as the missing Shapiro steps up to higher order, or some of the zero-bias peak regimes, the data fulfill several expected signatures, including secondary criteria. Yet, with sufficient depth of experimentation, backed by appropriate volumes of data, it should be possible to establish whether or not we are dealing with interesting new phenomena related to the topology of electron bands, quantum computing, and other topics within condensed matter physics.

In some situations, there is no independent reference for where the exotic state should reside in parameter space. In this case the only path forward is to search for a signal that matches the expectation. However, there is no guarantee that the signal found this way does not originate from a different effect. In these cases, it is particularly important to study the full phenomenology, explore the parameters as widely as possible within the experimental constraints, and try to figure out whether the apparent smoking gun can be a result of fine-tuning of trivial signals.

The most straightforward way to increase verifiability of a given experiment is to provide the community with full or ample data obtained over the course of measurements, from all samples studied within the project. The larger volume of information available helps pin down the plausible explanations. When more than one possibility is identified, it would be helpful to discuss the alternative and provide comprehensive arguments for and against each hypothesis, including why researchers favor one over the other. Sharing numerical data and code used for computation can also help. The primary burden is on the researchers making a claim to prove the existence of the more exotic state and to carefully exclude more conventional and more likely explanations. By sharing materials researchers invite the entire community to collaborate on their findings and strengthen and debug their conclusions.

Does the smoking gun approach to discovery still work? In principle yes, though true smoking gun single-plot proofs of breakthrough phenomena are certainly rare. Presenting a striking dataset can be helpful as a means of illustrating an interesting claim. However, the paper should then make it clear that the signal persists, and disclose how fine-tuned and selected the figures are, what the second-best regime or sample looks like, and how many high-quality samples and datasets the authors were able to obtain.

**Data availability**

Data are available through Zenodo at DOI: 10.5281/zenodo.8349309.

**Acknowledgements**

InSb nanowires are provided by E. Bakkers and G. Badawy. PbTe nanowires are provided by E. Bakkers and S. Schellingerhout. InAs/Al nanowires are provided by P. Krogstrup and S. Khan. Sn shells on nanowires and Al shells on 2DEGs were grown by C. Dempsey, M. Pendharkar and C. Palmstrøm, InAs quantum wells were grown by J.S. Lee, S. Harrington, B. Shojaei, J. Dong and C. Palmstrøm.

**Funding**

Work on identifying and analyzing smoking gun signatures is supported by the U.S. Department of Energy, Basic Energy Sciences grant DE-SC-0022073. Experiments on Sn-InSb nanowire and Al-InAs quantum well junctions are supported by NSF PIRE:HYBRID OISE-1743717, NSF Quantum Foundry funded via the Q-AMASE-i program under award DMR-1906325, U.S. ONR and ARO and France ANR through Grant No. ANR-17-PIRE-0001 (HYBRID), as well as through the France CNRS IRP HYNATOQ. Experiments on nanowire junctions with micromagnets and on PbTe nanowire quantum dots are supported by the U.S. Department of Energy Basic Energy Sciences under grant DE-SC-0019274.

## Supplementary Materials

### Methods

All measurements are done in dilution refrigerators with base temperatures ranging between 30 mK and 100 mK equipped with various cryogenic magnets. Standard low frequency lockin and DC measurements are performed. The RF radiation is applied via an antenna above the device. Attenuators (36 dB) are installed along the microwave line. Below we outline the sample fabrication steps.

Sn/InSb nanowire junctions. Josephson junctions are prepared by coating the standing InSb nanowires with a 15 nm layer of Sn(*25*). In front of the nanowire, another nanowire shadows the flux of Sn to create two disconnected Sn segments. Shadow junction wires are transferred onto chips patterned with local gates (1.5/6 nm Ti/PdAu covered by 10 nm of ALD HfOx), contacts to wires are made using standard electron beam lithography at 10 kV and thin film deposition of 10/140 nm Ti/Au.

InAs-Al nanowire device: InAs nanowires with an in-situ grown Al layer (*64*) are transferred onto a back-gate chip. A section of Al layer is etched by CD26:water (1:20). Ti/Au (10 nm/140 nm) is deposited on Al and bare InAs nanowire. Two CoFe strips are deposited close to the junction of Al/InAs and InAs to provide local magnetic fields.

Planar Hybrid Junctions. Josephson junctions are fabricated using the nanowire shadow mask method. The full method and material structures are described in Ref. (*46*). Typical values of the junction length and width are 100 nm and 5 μm, respectively. The Al film is deposited at cryogenic temperatures with a thickness of 10 nm.

PbTe nanowire quantum dots. The quantum dot devices are fabricated with PbTe nanowires. Contacts to the nanowires are made using electron beam lithography and thin film deposition of 10/160 nm Ti/Au. These samples are coated in 10 nm of $HfO_x$ deposited by ALD. A set of narrow gates is then deposited on the section of nanowire lying between the contacts. For stability and improved contact, the gates are deposited in a series of steps without breaking vacuum: first, 6/14 nm Ti/Au is deposited directly onto the nanowire, then 6/24 nm Ti/Au is deposited at angles of +/-45 degrees to the nanowire.

### Duration and Volume of Study

The Sn-InSb nanowires junction project lasts between 2018-06 to 2022-12. There are 61 devices across 9 chips that were fabricated and measured during 12 cooldowns in dilution refrigerators. About 5900 datasets were generated during this project.

We measured 19 InAs-Al nanowire tunneling devices from 3 separated dilution fridge cooldowns, acquiring more than 4000 data sets. We observed clear ABSs in 4 of the devices.

The fractional Josephson effect project lasts between 2021-03 to 2022-02. We measured 62 devices on 6 chips during 8 cooldowns in dilution refrigerators. About 5700 datasets are generated.

The PbTe nanowire quantum dot project started in 2022-01 and is ongoing at the time of writing. We have measured 27 devices across 9 chips during 10 cooldowns in dilution refrigerators. Approximately 1900 datasets have been generated during this project.

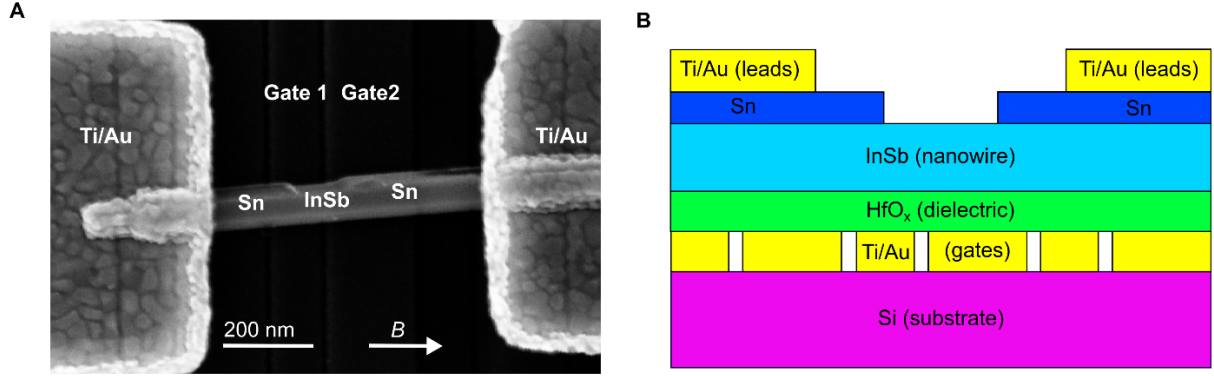

**Fig. S1. (A)** SEM picture of the nanowire Josephson junction studied in Fig. 1, the direction of external field is indicated with an arrow. **(B)** Cross-section schematic of the device. The nanowires growth method is introduced in Ref.(*25*). The nanowire is transferred onto a Silicon chip with pre-patterned bottom gates, using a micro-manipulator under an optical microscope. Detailed method and material information can be found in Refs. (*26, 27*).

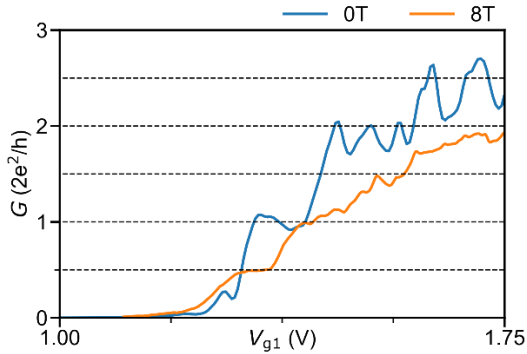

**Fig. S2.** Gate dependence of normal state differential conductance ($G$). The $G$ vs $V_{g1}$ traces are extracted from a 2D conductance map measured at $B$ = 0T and 8T.

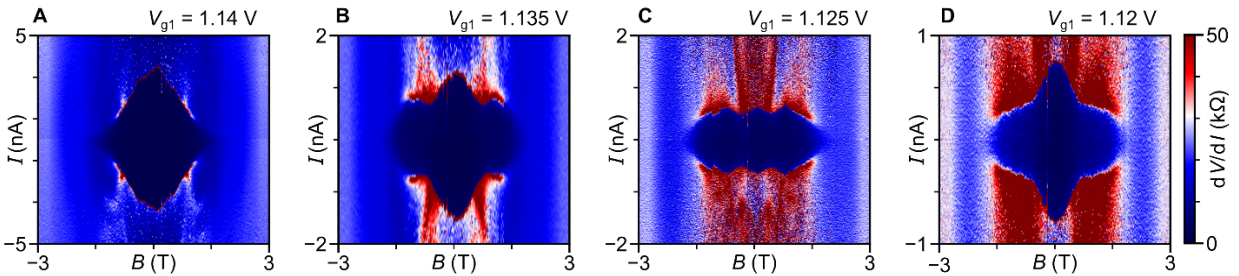

**Fig. S3. Supercurrent diffraction patterns at different $V_{g1}$. (A)** At $V_{g1}$ = 1.14 V, the magnitude of supercurrent decreases with the parallel field. **(B-D)** Increase supercurrent appear at $V_{g1}$ = 1.135-1.125 V, disappear at $V_{g1}$ = 1.12 V. All the diffraction patterns are measured at $V_{g2}$ = -2.5 V. All the scans are done by sweeping from zero current bias to the positive and negative bias separately. More supercurrent diffraction patterns and discussion are presented in Ref. (*26*).

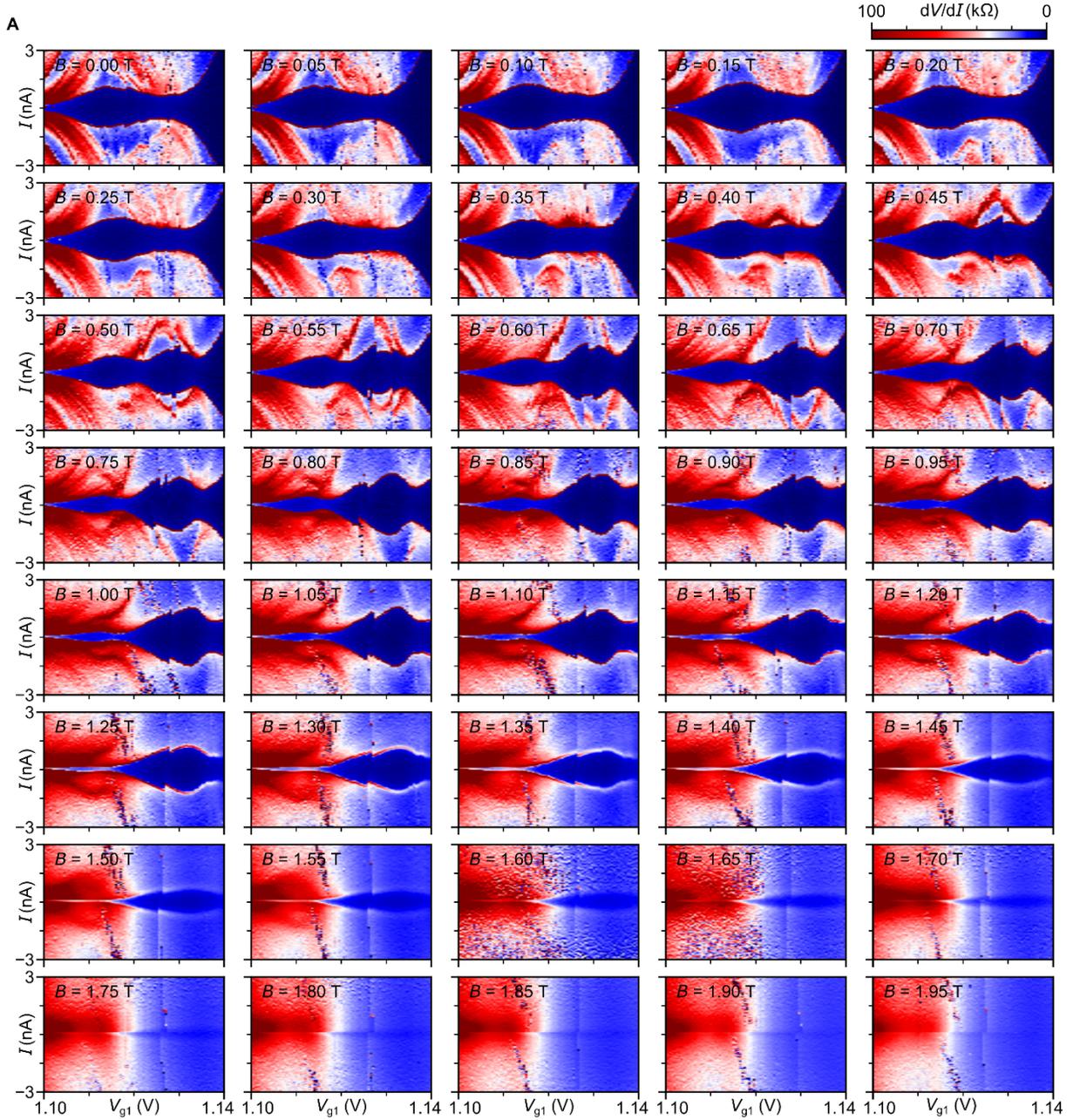

**Fig. S4. Gate voltage dependence of increased supercurrent in parallel field.** Additional data of Fig. 1(C). The parallel field is ramped up with a step of 0.05 T. A strong resonance appears at $B$ = 0.4 T between $V_g$ = 1.12- 1.14 V and moving to larger bias, followed by the increased Josephson current in the field. All the measurements are done at $V_{g2}$ = -2.5 V. All the scans are done by sweeping from zero current bias to the positive and negative bias separately. More supercurrent diffraction patterns and discussion are presented in Ref. (*26*).

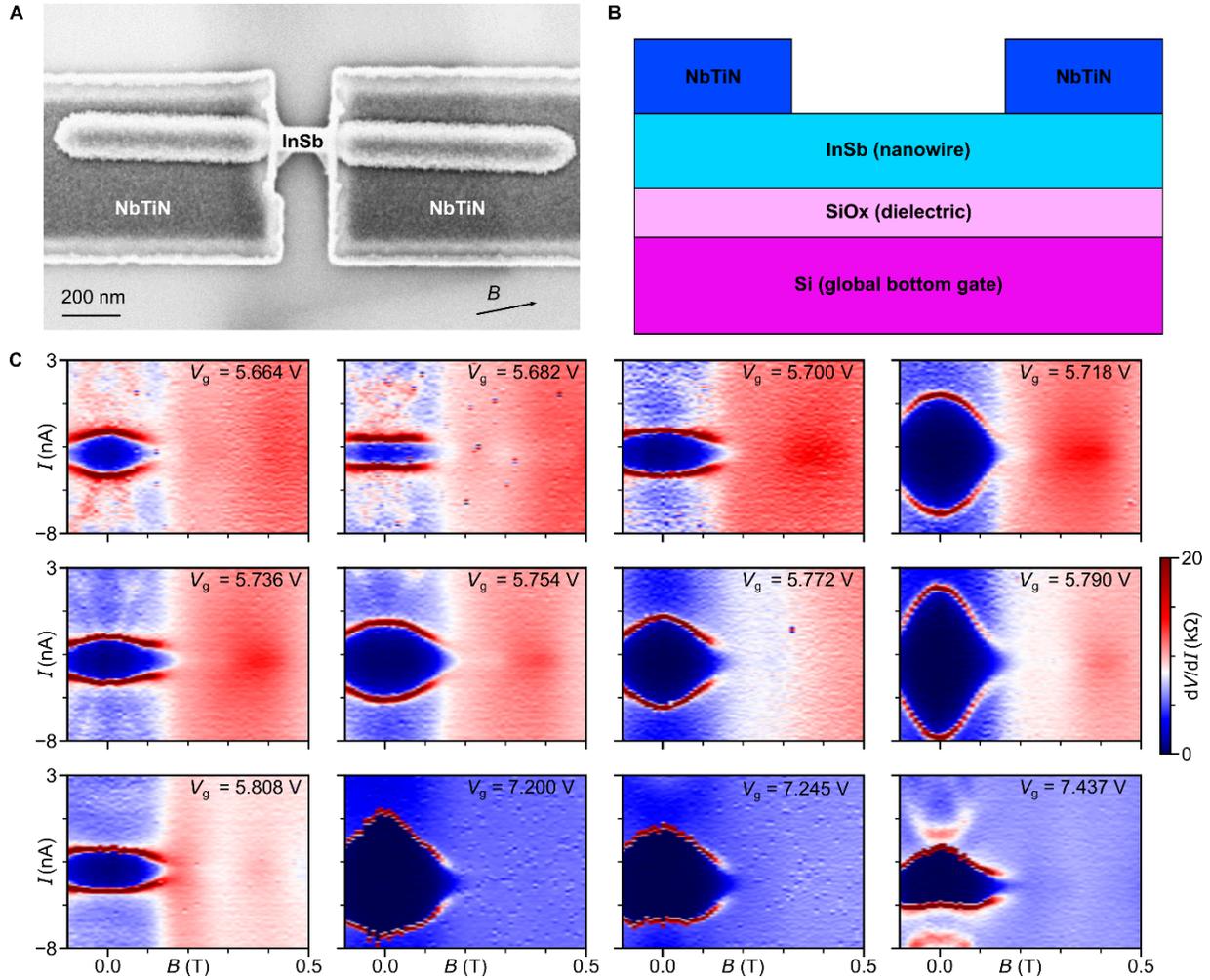

**Fig. S5. Supercurrent diffraction patterns at different $V_g$, measured in an InSb nanowire with NbTiN contacts (A)** SEM picture of nanowire Josephson junction made with InSb nanowires and NbTiN superconductor leads created by sputter deposition after sulfur passivation and sputter cleaning of the InSb surface. **(B)** Cross-section schematic of the device. **(C)** Supercurrent diffraction pattern measured at several different gate voltages. All the scans are done by sweeping from negative bias to the positive bias. There is an offset of ~-3 nA present in all raw data and it is due to the measurement setup. During data processing, the y-axis is shifted by 3 nA to reduce the effect. Detailed fabrication method, material information, and discussion are presented in the supplementary of Ref. (*65*).

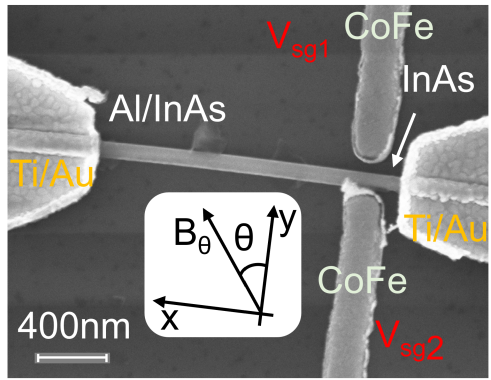

**Fig. S6**. InAs/Al device image with CoFe micromagnet gates. The coordinate system demonstrates the direction of the applied magnetic field.

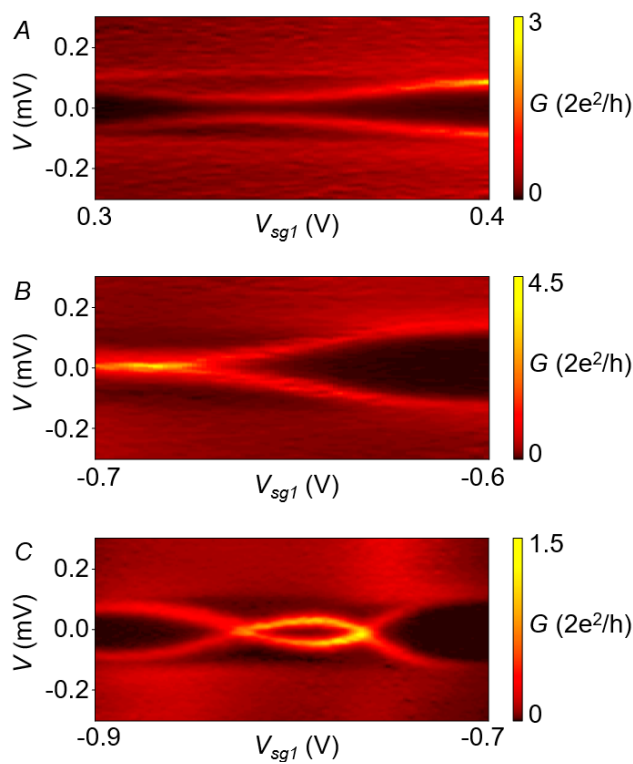

**Fig. S7 Examples of Andreev Bound States spectra from the same device.** **(A)** $V_g$ = 6.075 V. There is no zero-bias peak. **(B)** $V_g$ = 6.15V. There is a zero-bias peak and the peak sticks for some range of $V_{sg1}$. The peak is not a plateau, and its magnitude can be larger than $2e^2/h$. **(C)** $V_g$ = 6.45V. There is a zero-bias peak, but the peak does not stick to zero. $V_{sg2}$ = 0 V. External magnetic field is 0. CoFe strips are magnetized.

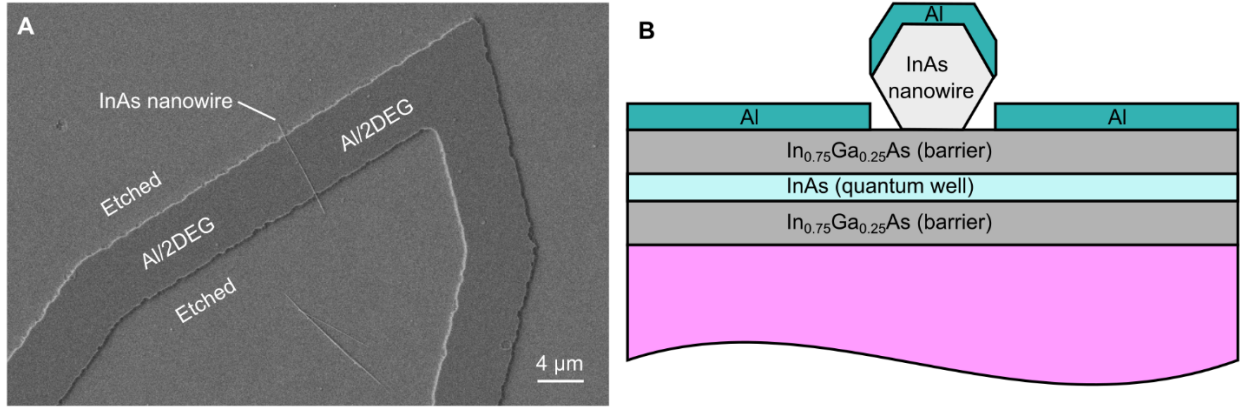

**Fig. S8. (A)** SEM picture of the planar Josephson junction studied in Fig. 3. **(B)** Cross-section schematic of the device. The nanowire is transferred randomly from a nanowire mother chip and is used as a shadow mask during Al deposition (*47*). Detailed method and material information can be found in Ref. (*46*).

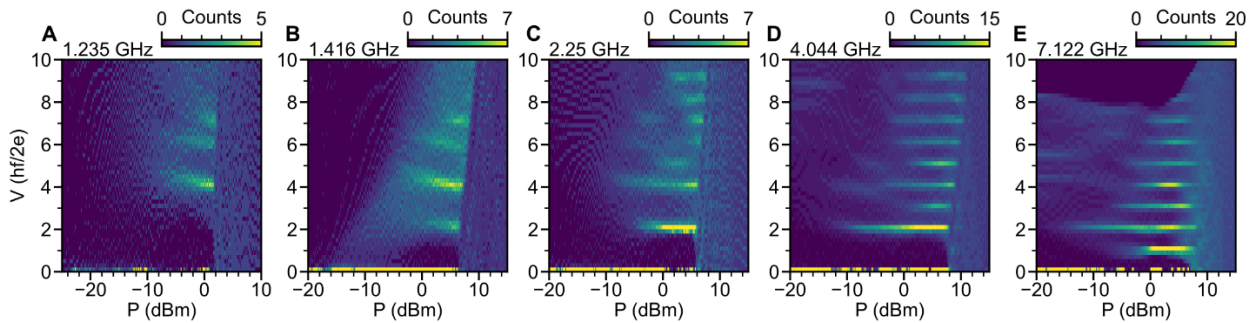

**Fig. S9. Frequency dependence of missing Shapiro steps. (A)** At 1.235 GHz, not only odd steps (V = 1, 3, 5) but also an even step (V = 2) is missing. **(B)** Duplicated figure of Fig. 3C showing a series of missing odd steps up to V = 5. **(C-E)** Missing steps reappear as the frequency increases. Suppressed steps are observed, e.g., in panel (C) between V = 5 and V = 8. The microwave radiation frequency is noted in each panel. More frequency-dependent data and discussion is presented in Ref. (*47*).

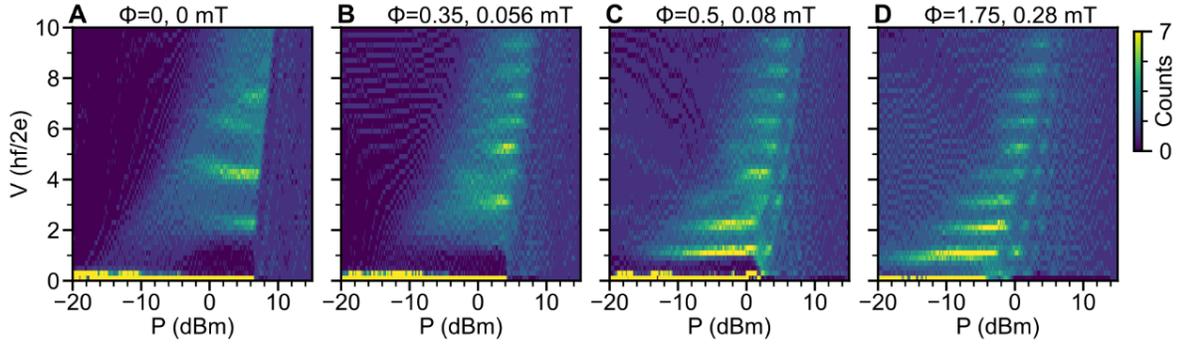

**Figure S10. Missing Shapiro steps at several magnetic fields.** Φ is the calculated magnetic flux quantum penetrating the junction. The microwave radiation frequency is 1.416 GHz. **(A)** The same regime (from a different dataset) as that in Fig. 3C, a series of missing odd steps up to $V = 5$ are observed. **(B)** At Φ = 0.35, two odd steps ($V = 3, 5$) appears while two even steps ($V = 2, 4$) are suppressed. **(C)** At Φ = 0.5, first and second steps are pronounced while the third step is strongly suppressed. **(D)** At Φ = 1.75, all steps are visible. A quantitative model describing the behavior of missing steps in magnetic field is yet to be found. An offset of 0.213 mT is applied numerically to the magnetic field to make the critical current maximize at zero field. The offset may be due to flux trapped in devices or in the magnet. More field-dependent data and discussion is presented in Ref. (*47*).

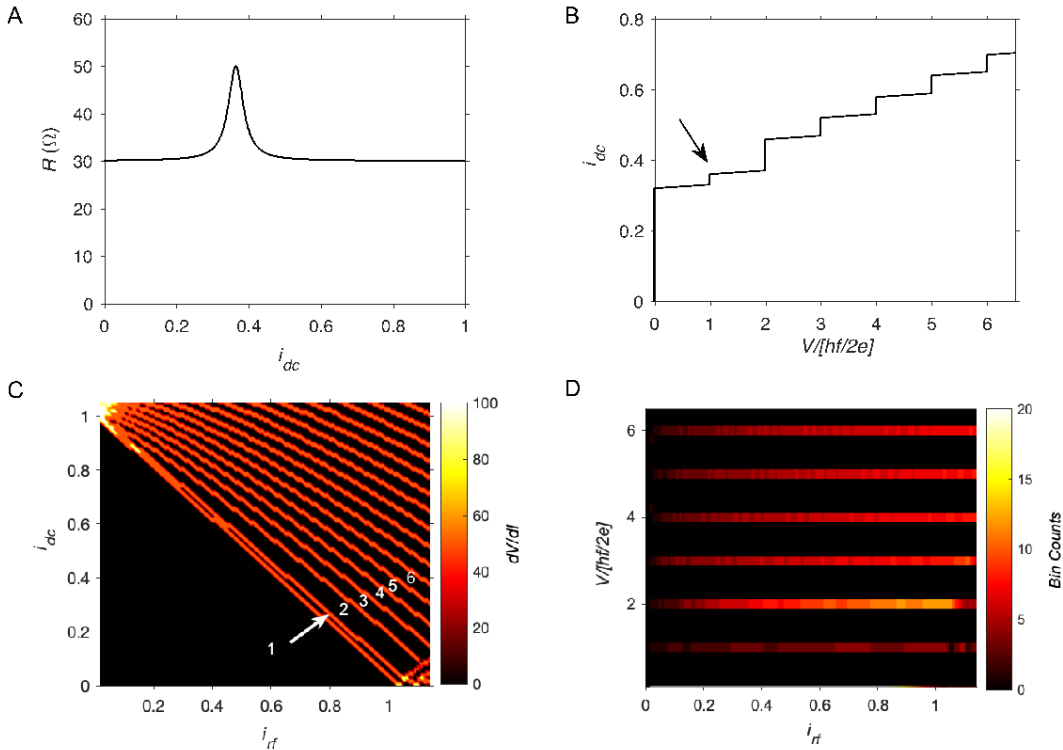

**Fig. S11. Simulated suppressed Shapiro steps due to bias-dependent resistance.** **(A)** $R(i_{dc})$ as a function of current bias $i_{dc}$ at rf drive amplitude $i_{rf} = 0.7$, **(B)** $i_{dc}$ vs. Shapiro step number $n = V/[hf/2e]$ at $i_{rf} = 0.7$ showing the suppression of first Shapiro step, **(C)** $dV/dI$ as a function of $i_{dc}$ and $i_{rf}$ showing suppression of first Shapiro step over the entire range of $i_{rf}$, **(D)** Histogram plot corresponding to (C). Lower bin counts for step 1 indicate a suppressed step.

**Shapiro Step Numerical Simulation Details:**

In this section, we look at the effect of non-linear resistance on the suppression of Shapiro steps. The plots presented in Fig. S11 are based on the phenomenological model presented in Ref. (*66*). We show that in the presence of a bias dependent resistance, it is possible to suppress one or more Shapiro steps even in the absence of a 4π – periodic supercurrent, i.e. a 4π – periodic supercurrent need not unambiguously identify Majorana modes in Shapiro step experiments.

We define a bias dependent resistance $R(i_{dc})$ using a Lorentzian peak shape of HWHM = 0.025 and peak height of 20 Ω superimposed on a constant resistance background of 30 Ω :

$$R(i_{dc}) = 30 + 20 * \frac{(0.025)^2}{(0.025)^2 + (i_{dc} - s_i)^2} \tag{1}$$

Where $i_{dc} = I_{dc}/I_c$ and $s_i$ is the bias current corresponding to the $i^{th}$ Shapiro step. We use this $R(i_{dc})$ in the first order non-linear differential equation obtained from the first and second Josephson relations :

$$\dot{\phi} = \frac{2eR(i_{dc})I_c}{\frac{h}{2\pi}} [i_{rf} \sin(2\pi ft) + i_{dc} - \sin(\phi)] \tag{2}$$

Here, ϕ is the superconducting phase difference, e is the charge of an electron, Ic is the supercurrent (3.3 µA) at zero rf-power, f = 2 GHz is the frequency and $i_{rf} = I_{rf}/I_c$. We have used the RK4 method to solve this non-linear differential equation.

Fig S11(A) shows the form of the resistance peak as a function of $i_{dc}$ as defined in equation (1). In Fig S11(B), we show a line-cut at $i_{rf}$ = 0.7 with the resistance peak placed at the first Shapiro step. We see that the first Shapiro step is suppressed while the second step is enlarged. In Fig S11(C), we show a plot of differential resistance as a function of $i_{dc}$ and $i_{rf}$. For this plot we make the resistance peak coincide with the first Shapiro step at each $i_{rf}$ using the method described in Ref. (*66*). We see that the first step can be suppressed for the entire range of $i_{rf}$. We can also present the data in Fig S11(C) in the form of a histogram where the voltage is binned and we count the number of points in each voltage bin (Fig.S11(D)). Fewer bin counts represent a suppressed step.

We can also suppress multiple Shapiro steps using this method, including even steps as illustrated in Ref. (*66*).

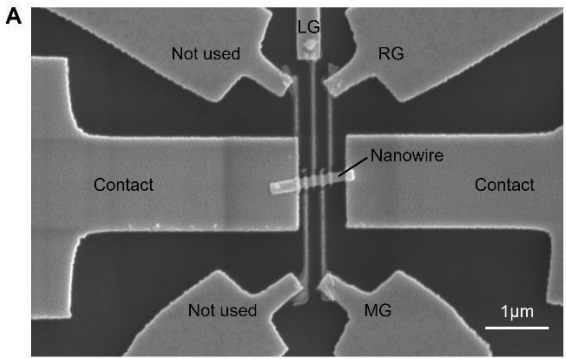
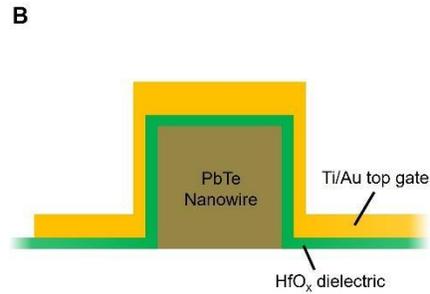

**Fig. S12. PbTe nanowire quantum dot device. (A)** SEM image of the nanowire quantum dot device studied in Fig. 4. **(B)** Schematic diagram of gates deposited on nanowire with a thin dielectric layer separating them. See Methods for a summary of the fabrication process.

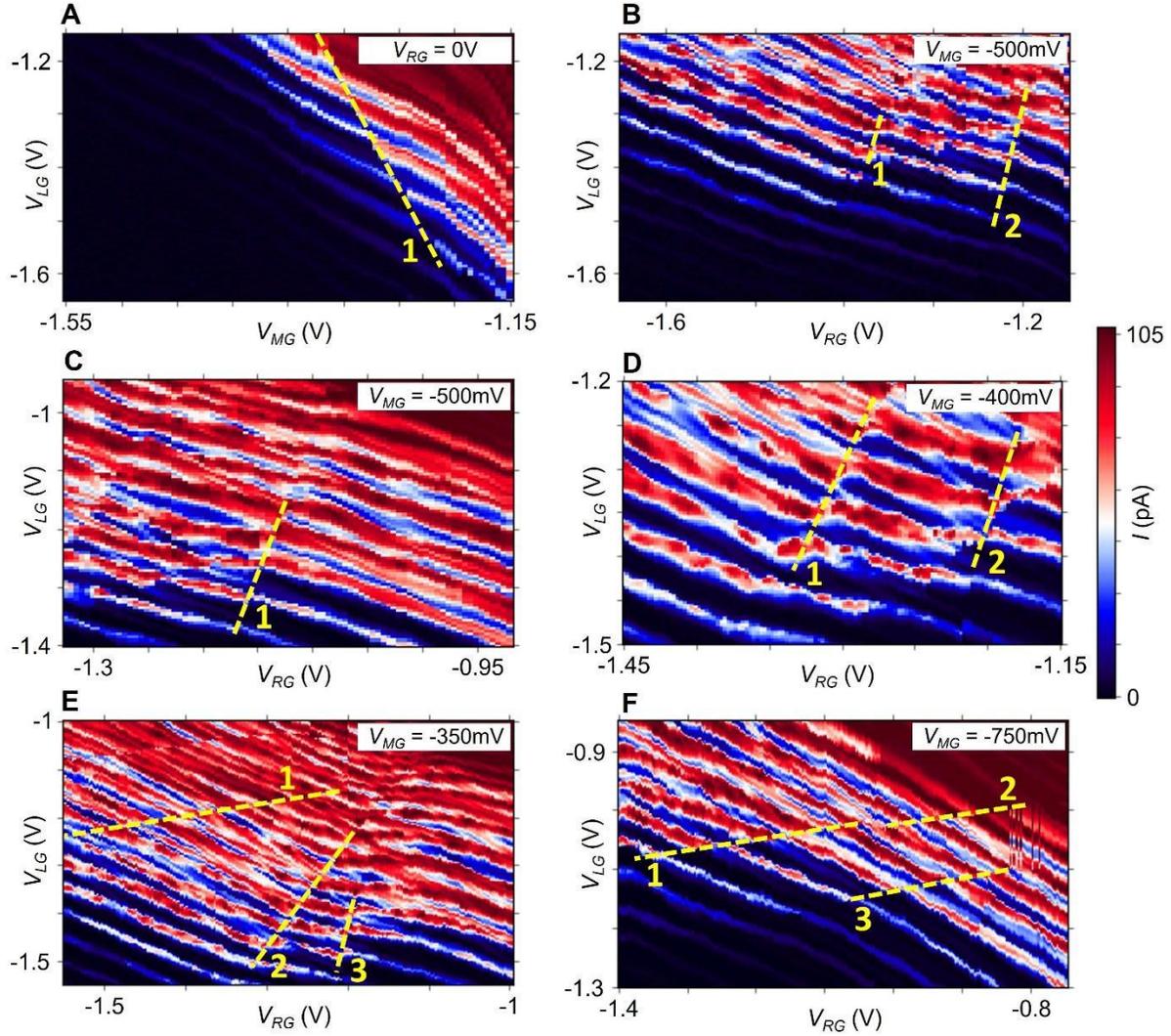

| Subfigure | $\left(\frac{\Delta\theta}{2\pi}\right)_1$ | $\left(\frac{\Delta\theta}{2\pi}\right)_2$ | $\left(\frac{\Delta\theta}{2\pi}\right)_3$ |
|---|---|---|---|
| A | +0.21 ± 0.03 | - | - |
| B | -0.38 ± 0.05 | -0.27 ± 0.04 | - |
| C | -0.19 ± 0.03 | - | - |
| D | -0.38 ± 0.04 | -0.47 ± 0.07 | - |
| E | -0.61 ± 0.12 | -0.57 ± 0.08 | -0.30 ± 0.02 |
| F | -0.40 ± 0.06 | -0.69 ± 0.02 | -0.28 ± 0.02 |

**Fig. S13. Datasets with diagonal jump shifts in PbTe nanowire quantum dots. (A-F)** Further examples of the gate voltage dependence of conductance, with yellow dashed lines indicating pattern skips. **(G)** Table summarizing the magnitudes of pattern shifts. $(\Delta\theta/2\pi)_n$ is defined to be the average

magnitude of resonance shifting across line *n*, expressed as a fraction of their period. The period of a set of resonances is taken to be the average distance between each pair of adjacent resonances, as measured parallel to the observed diagonal shifting line. The magnitude of each shift is found similarly as the diagonally measured distance between a resonance's midpoint before and after a shift.

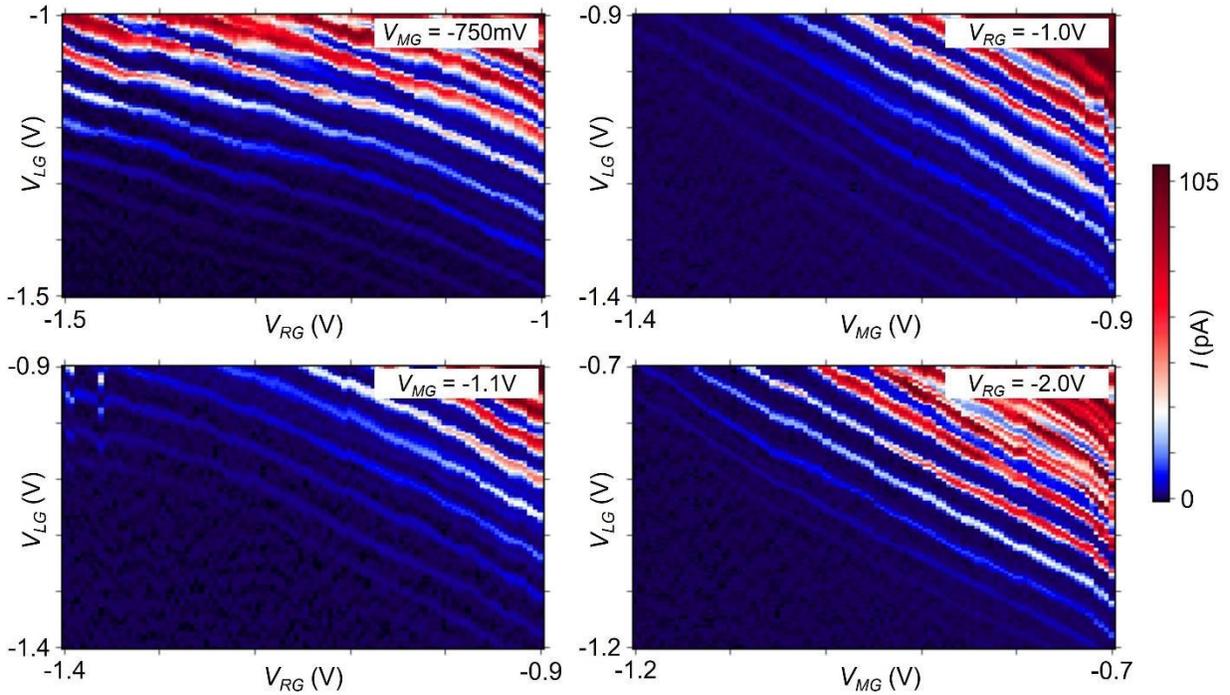

**Fig. S14. Datasets without diagonal jump shifts in PbTe nanowire quantum dots.** Examples of typical datasets showing gate voltage dependence of conductance. Resonances are not observed to shift in these regimes outside of small shifts due to noise. Further tuning, or wider sweeps, of gate voltages is necessary in such cases to identify jump shifts.

# References

1. M. Z. Hasan, C. L. Kane, Colloquium: Topological insulators. *Rev. Mod. Phys.* **82**, 3045–3067 (2010).

2. G. Moore, N. Read, Nonabelions in the Fractional Quantum Hall-Effect. *Nucl. Phys. B*. **360**, 362–396 (1991).

3. A. Stern, N. H. Lindner, Topological Quantum Computation-From Basic Concepts to First Experiments. *Science*. **339**, 1179–1184 (2013).

4. K. v. Klitzing, G. Dorda, M. Pepper, New Method for High-Accuracy Determination of the Fine-Structure Constant Based on Quantized Hall Resistance. *Phys. Rev. Lett.* **45**, 494–497 (1980).

5. D. A. Wollman, D. J. Van Harlingen, W. C. Lee, D. M. Ginsberg, A. J. Leggett, Experimental determination of the superconducting pairing state in YBCO from the phase coherence of YBCO-Pb dc SQUIDs. *Phys. Rev. Lett.* **71**, 2134–2137 (1993).

6. A. Wallraff, D. I. Schuster, A. Blais, L. Frunzio, R.-S. Huang, J. Majer, S. Kumar, S. M. Girvin, R. J. Schoelkopf, Strong coupling of a single photon to a superconducting qubit using circuit quantum electrodynamics. *Nature*. **431**, 162–167 (2004).

7. D. Hsieh, D. Qian, L. Wray, Y. Xia, Y. S. Hor, R. J. Cava, M. Z. Hasan, A topological Dirac insulator in a quantum spin Hall phase. *Nature*. **452**, 970-U5 (2008).

8. D. Aasen, M. Hell, R. V. Mishmash, A. Higginbotham, J. Danon, M. Leijnse, T. S. Jespersen, J. A. Folk, C. M. Marcus, K. Flensberg, J. Alicea, Milestones Toward Majorana-Based Quantum Computing. *Phys. Rev. X*. **6**, 031016 (2016).

9. C. J. Trimble, M. T. Wei, N. F. Q. Yuan, S. S. Kalantre, P. Liu, H.-J. Han, M.-G. Han, Y. Zhu, J. J. Cha, L. Fu, J. R. Williams, Josephson detection of time-reversal symmetry broken superconductivity in SnTe nanowires. *Npj Quantum Mater.* **6**, 1–6 (2021).

10. Y. Oreg, G. Refael, F. von Oppen, Helical Liquids and Majorana Bound States in Quantum Wires. *Phys. Rev. Lett.* **105**, 177002 (2010).

11. R. M. Lutchyn, J. D. Sau, S. Das Sarma, Majorana Fermions and a Topological Phase Transition in Semiconductor-Superconductor Heterostructures. *Phys. Rev. Lett.* **105**, 077001 (2010).

12. M. Wimmer, A. R. Akhmerov, J. P. Dahlhaus, C. W. J. Beenakker, Quantum point contact as a probe of a topological superconductor. *New J. Phys.* **13** (2011), doi:Artn 053016 Doi 10.1088/1367-2630/13/5/053016.

13. E. Bocquillon, R. S. Deacon, J. Wiedenmann, P. Leubner, T. M. Klapwijk, C. Brüne, K. Ishibashi, H. Buhmann, L. W. Molenkamp, Gapless Andreev bound states in the quantum spin Hall insulator HgTe. *Nat. Nanotechnol.* **12**, 137–143 (2017).

14. D. E. Feldman, B. I. Halperin, Fractional charge and fractional statistics in the quantum Hall effects. *Rep. Prog. Phys.* **84**, 076501 (2021).

15. S. Lee, J.-H. Kim, Y.-W. Kwon, The First Room-Temperature Ambient-Pressure Superconductor (2023), , doi:10.48550/arXiv.2307.12008.

16. K. Guo, Y. Li, S. Jia, Ferromagnetic half levitation of LK-99-like synthetic samples. *Sci. China Phys. Mech. Astron.* **66**, 107411 (2023).

17. Y. Cao, V. Fatemi, S. Fang, K. Watanabe, T. Taniguchi, E. Kaxiras, P. Jarillo-Herrero, Unconventional superconductivity in magic-angle graphene superlattices. *Nature*. **556**, 43–50 (2018).

18. S. Sasaki, M. Kriener, K. Segawa, K. Yada, Y. Tanaka, M. Sato, Y. Ando, Topological Superconductivity in CuxBi2Se3. *Phys. Rev. Lett.* **107**, 217001 (2011).

19. M. Popinciuc, V. E. Calado, X. L. Liu, A. R. Akhmerov, T. M. Klapwijk, L. M. K. Vandersypen, Zero-bias conductance peak and Josephson effect in graphene-NbTiN junctions. *Phys. Rev. B*. **85**, 205404 (2012).

20. Y. Saito, J. Ge, K. Watanabe, T. Taniguchi, A. F. Young, Independent superconductors and correlated insulators in twisted bilayer graphene. *Nat. Phys.* **16**, 926–930 (2020).

21. S. Ran, C. Eckberg, Q.-P. Ding, Y. Furukawa, T. Metz, S. R. Saha, I.-L. Liu, M. Zic, H. Kim, J. Paglione, N. P. Butch, Nearly ferromagnetic spin-triplet superconductivity. *Science*. **365**, 684–687 (2019).

22. H. Zhou, L. Holleis, Y. Saito, L. Cohen, W. Huynh, C. L. Patterson, F. Yang, T. Taniguchi, K. Watanabe, A. F. Young, Isospin magnetism and spin-polarized superconductivity in Bernal bilayer graphene. *Science*. **375**, 774–778 (2022).

23. J. Jang, D. G. Ferguson, V. Vakaryuk, R. Budakian, S. B. Chung, P. M. Goldbart, Y. Maeno, Observation of Half-Height Magnetization Steps in Sr2RuO4. *Science*. **331**, 186–188 (2011).

24. B. A. Bernevig, "Topological Insulators and Topological Superconductors" in *Topological Insulators and Topological Superconductors* (Princeton University Press, 2013; https://www.degruyter.com/document/doi/10.1515/9781400846733/html).

25. M. Pendharkar, B. Zhang, H. Wu, A. Zarassi, P. Zhang, C. P. Dempsey, J. S. Lee, S. D. Harrington, G. Badawy, S. Gazibegovic, R. L. M. Op het Veld, M. Rossi, J. Jung, A.-H. Chen, M. A. Verheijen, M. Hocevar, E. P. A. M. Bakkers, C. J. Palmstrøm, S. M. Frolov, Parity-preserving and magnetic field–resilient superconductivity in InSb nanowires with Sn shells. *Science*. **372**, 508–511 (2021).

26. B. Zhang, Z. Li, H. Wu, M. Pendharkar, C. Dempsey, J. S. Lee, S. D. Harrington, C. J. Palmstrom, S. M. Frolov, Supercurrent through a single transverse mode in nanowire Josephson junctions (2023), , doi:10.48550/arXiv.2306.00146.

27. B. Zhang, Z. Li, V. Aguilar, P. Zhang, M. Pendharkar, C. Dempsey, J. S. Lee, S. D. Harrington, S. Tan, J. S. Meyer, M. Houzet, C. J. Palmstrom, S. M. Frolov, Evidence of $\phi$0-Josephson junction from skewed diffraction patterns in Sn-InSb nanowires (2023), , doi:10.48550/arXiv.2212.00199.

28. S. M. Frolov, M. J. Manfra, J. D. Sau, Topological superconductivity in hybrid devices. *Nat. Phys.* **16**, 718–724 (2020).

29. Y. Sato, K. Ueda, Y. Takeshige, H. Kamata, K. Li, L. Samuelson, H. Q. Xu, S. Matsuo, S. Tarucha, Quasiparticle Trapping at Vortices Producing Josephson Supercurrent Enhancement. *Phys. Rev. Lett.* **128**, 207001 (2022).

30. A. Rogachev, T.-C. Wei, D. Pekker, A. T. Bollinger, P. M. Goldbart, A. Bezryadin, Magnetic-Field Enhancement of Superconductivity in Ultranarrow Wires. *Phys. Rev. Lett.* **97**, 137001 (2006).

31. Y. Chen, Y.-H. Lin, S. D. Snyder, A. M. Goldman, A. Kamenev, Dissipative superconducting state of non-equilibrium nanowires. *Nat. Phys.* **10**, 567–571 (2014).

32. V. Mourik, K. Zuo, S. M. Frolov, S. R. Plissard, E. P. A. M. Bakkers, L. P. Kouwenhoven, Signatures of Majorana Fermions in Hybrid Superconductor-Semiconductor Nanowire Devices. *Science*. **336**, 1003–1007 (2012).

33. E. J. H. Lee, X. Jiang, R. Aguado, G. Katsaros, C. M. Lieber, S. De Franceschi, Zero-bias anomaly in a nanowire quantum dot coupled to superconductors. *Phys. Rev. Lett.* **109**, 186802 (2012).

34. E. J. H. Lee, X. Jiang, M. Houzet, R. Aguado, C. M. Lieber, S. De Franceschi, Spin-resolved Andreev levels and parity crossings in hybrid superconductor–semiconductor nanostructures. *Nat. Nanotechnol.* **9**, 79–84 (2013).

35. Y. Jiang, M. Gupta, C. Riggert, M. Pendharkar, C. Dempsey, J. S. Lee, S. D. Harrington, C. J. Palmstrøm, V. S. Pribiag, S. M. Frolov, Zero-bias conductance peaks at zero applied magnetic field due to stray fields from integrated micromagnets in hybrid nanowire quantum dots (2023), , doi:10.48550/arXiv.2305.19970.

36. S. Vaitiekėnas, Y. Liu, P. Krogstrup, C. M. Marcus, Zero-bias peaks at zero magnetic field in ferromagnetic hybrid nanowires. *Nat. Phys.* **17**, 43–47 (2021).

37. M. Jardine, J. Stenger, Y. Jiang, E. J. de Jong, W. Wang, A. C. Bleszynski Jayich, S. M. Frolov, Integrating micromagnets and hybrid nanowires for topological quantum computing. *SciPost Phys.* **11**, 090 (2021).

38. P. Yu, J. Chen, M. Gomanko, G. Badawy, E. P. a. M. Bakkers, K. Zuo, V. Mourik, S. M. Frolov, Non-Majorana states yield nearly quantized conductance in proximatized nanowires. *Nat. Phys.* **17**, 482–488 (2021).

39. Z. Wang, H. Song, D. Pan, Z. Zhang, W. Miao, R. Li, Z. Cao, G. Zhang, L. Liu, L. Wen, R. Zhuo, D. E. Liu, K. He, R. Shang, J. Zhao, H. Zhang, Plateau Regions for Zero-Bias Peaks within 5% of the Quantized Conductance Value $2{e}^{2}/h$. *Phys. Rev. Lett.* **129**, 167702 (2022).

40. Microsoft Quantum, M. Aghaee, A. Akkala, Z. Alam, R. Ali, A. Alcaraz Ramirez, M. Andrzejczuk, A. E. Antipov, P. Aseev, M. Astafev, B. Bauer, J. Becker, S. Boddapati, F. Boekhout, J. Bommer, T. Bosma, L. Bourdet, S. Boutin, P. Caroff, L. Casparis, M. Cassidy, S. Chatoor, A. W. Christensen, N. Clay, W. S. Cole, F. Corsetti, A. Cui, P. Dalampiras, A. Dokania, G. de Lange, M. de Moor, J. C. Estrada Saldaña, S. Fallahi, Z. H. Fathabad, J. Gamble, G. Gardner, D. Govender, F. Griggio, R. Grigoryan, S. Gronin, J. Gukelberger, E. B. Hansen, S. Heedt, J. Herranz Zamorano, S. Ho, U. L. Holgaard, H. Ingerslev, L. Johansson, J. Jones, R. Kallaher, F. Karimi, T. Karzig, C. King, M. E. Kloster, C. Knapp, D. Kocon, J. Koski, P. Kostamo, P. Krogstrup, M. Kumar, T. Laeven, T. Larsen, K. Li, T. Lindemann, J. Love, R. Lutchyn, M. H. Madsen, M. Manfra, S. Markussen, E. Martinez, R. McNeil, E. Memisevic, T. Morgan, A. Mullally, C. Nayak, J. Nielsen, W. H. P. Nielsen, B.

Nijholt, A. Nurmohamed, E. O'Farrell, K. Otani, S. Pauka, K. Petersson, L. Petit, D. I. Pikulin, F. Preiss, M. Quintero-Perez, M. Rajpalke, K. Rasmussen, D. Razmadze, O. Reentila, D. Reilly, R. Rouse, I. Sadovskyy, L. Sainiemi, S. Schreppler, V. Sidorkin, A. Singh, S. Singh, S. Sinha, P. Sohr, T. Stankevič, L. Stek, H. Suominen, J. Suter, V. Svidenko, S. Teicher, M. Temuerhan, N. Thiyagarajah, R. Tholapi, M. Thomas, E. Toomey, S. Upadhyay, I. Urban, S. Vaitiekėnas, K. Van Hoogdalem, D. Van Woerkom, D. V. Viazmitinov, D. Vogel, S. Waddy, J. Watson, J. Weston, G. W. Winkler, C. K. Yang, S. Yau, D. Yi, E. Yucelen, A. Webster, R. Zeisel, R. Zhao, InAs-Al hybrid devices passing the topological gap protocol. *Phys. Rev. B*. **107**, 245423 (2023).

41. R. Hess, H. F. Legg, D. Loss, J. Klinovaja, Trivial Andreev Band Mimicking Topological Bulk Gap Reopening in the Nonlocal Conductance of Long Rashba Nanowires. *Phys. Rev. Lett.* **130**, 207001 (2023).

42. D. Pekker, C.-Y. Hou, V. E. Manucharyan, E. Demler, Proposal for Coherent Coupling of Majorana Zero Modes and Superconducting Qubits Using the $4\pi$ Josephson Effect. *Phys. Rev. Lett.* **111**, 107007 (2013).

43. L. P. Rokhinson, X. Liu, J. K. Furdyna, Observation of the fractional ac Josephson effect: the signature of Majorana particles (2012), doi:10.1038/nphys2429.

44. J. Wiedenmann, E. Bocquillon, R. S. Deacon, S. Hartinger, O. Herrmann, T. M. Klapwijk, L. Maier, C. Ames, C. Brüne, C. Gould, A. Oiwa, K. Ishibashi, S. Tarucha, H. Buhmann, L. W. Molenkamp, 4π-periodic Josephson supercurrent in HgTe-based topological Josephson junctions. *Nat. Commun.* **7**, 10303 (2016).

45. F. Domínguez, F. Hassler, G. Platero, Dynamical detection of Majorana fermions in current-biased nanowires. *Phys. Rev. B*. **86**, 140503 (2012).

46. P. Zhang, A. Zarassi, M. Pendharkar, J. S. Lee, L. Jarjat, V. Van de Sande, B. Zhang, S. Mudi, H. Wu, S. Tan, C. P. Dempsey, A. P. McFadden, S. D. Harrington, B. Shojaei, J. T. Dong, A.-H. Chen, M. Hocevar, C. J. Palmstrøm, S. M. Frolov, Planar Josephson Junctions Templated by Nanowire Shadowing (2022), , doi:10.48550/arXiv.2211.04130.

47. P. Zhang, S. Mudi, M. Pendharkar, J. S. Lee, C. P. Dempsey, A. P. McFadden, S. D. Harrington, J. T. Dong, H. Wu, A.-H. Chen, M. Hocevar, C. J. Palmstrøm, S. M. Frolov, Missing odd-order Shapiro steps do not uniquely indicate fractional Josephson effect (2022), , doi:10.48550/arXiv.2211.08710.

48. V. S. Pribiag, A. J. A. Beukman, F. Qu, M. C. Cassidy, C. Charpentier, W. Wegscheider, L. P. Kouwenhoven, Edge-mode superconductivity in a two-dimensional topological insulator. *Nat. Nanotechnol.* **10**, 593–597 (2015).

49. A. De Cecco, K. Le Calvez, B. Sacépé, C. B. Winkelmann, H. Courtois, Interplay between electron overheating and ac Josephson effect. *Phys. Rev. B*. **93**, 180505 (2016).

50. M. C. Dartiailh, J. J. Cuozzo, B. H. Elfeky, W. Mayer, J. Yuan, K. S. Wickramasinghe, E. Rossi, J. Shabani, Missing Shapiro steps in topologically trivial Josephson junction on InAs quantum well. *Nat. Commun.* **12**, 78 (2021).

51. A. Stern, Anyons and the quantum Hall effect—A pedagogical review. *Ann. Phys.* **323**, 204–249 (2008).

52. H. L. Stormer, D. C. Tsui, A. C. Gossard, The fractional quantum Hall effect. *Rev. Mod. Phys.* **71**, S298–S305 (1999).

53. J. Nakamura, S. Liang, G. C. Gardner, M. J. Manfra, Direct observation of anyonic braiding statistics. *Nat. Phys.* **16**, 931–936 (2020).

54. A. Bid, N. Ofek, M. Heiblum, V. Umansky, D. Mahalu, Shot Noise and Charge at the $2/3$ Composite Fractional Quantum Hall State. *Phys. Rev. Lett.* **103**, 236802 (2009).

55. H. Bartolomei, M. Kumar, R. Bisognin, A. Marguerite, J.-M. Berroir, E. Bocquillon, B. Plaçais, A. Cavanna, Q. Dong, U. Gennser, Y. Jin, G. Fève, Fractional statistics in anyon collisions. *Science*. **368**, 173–177 (2020).

56. R. L. Willett, L. N. Pfeiffer, K. W. West, Measurement of filling factor 5/2 quasiparticle interference with observation of charge e/4 and e/2 period oscillations. *Proc. Natl. Acad. Sci.* **106**, 8853–8858 (2009).

57. M. Gomanko, E. J. de Jong, Y. Jiang, S. G. Schellingerhout, E. Bakkers, S. M. Frolov, Spin and Orbital Spectroscopy in the Absence of Coulomb Blockade in Lead Telluride Nanowire Quantum Dots. *SciPost Phys.* **13**, 089 (2022).

58. C. Barthel, M. Kjærgaard, J. Medford, M. Stopa, C. M. Marcus, M. P. Hanson, A. C. Gossard, Fast sensing of double-dot charge arrangement and spin state with a radio-frequency sensor quantum dot. *Phys. Rev. B*. **81**, 161308 (2010).

59. M. Pioro-Ladrière, J. H. Davies, A. R. Long, A. S. Sachrajda, L. Gaudreau, P. Zawadzki, J. Lapointe, J. Gupta, Z. Wasilewski, S. Studenikin, Origin of switching noise in GaAs∕AlxGa1-xAs lateral gated devices. *Phys. Rev. B*. **72**, 115331 (2005).

60. S. Vaitiekėnas, G. W. Winkler, B. van Heck, T. Karzig, M.-T. Deng, K. Flensberg, L. I. Glazman, C. Nayak, P. Krogstrup, R. M. Lutchyn, C. M. Marcus, Flux-induced topological superconductivity in full-shell nanowires. *Science*. **367**, eaav3392 (2020).

61. M. Valentini, F. Peñaranda, A. Hofmann, M. Brauns, R. Hauschild, P. Krogstrup, P. San-Jose, E. Prada, R. Aguado, G. Katsaros, Nontopological zero-bias peaks in full-shell nanowires induced by flux-tunable Andreev states. *Science*. **373**, 82–88 (2021).

62. M. Valentini, M. Borovkov, E. Prada, S. Martí-Sánchez, M. Botifoll, A. Hofmann, J. Arbiol, R. Aguado, P. San-Jose, G. Katsaros, Majorana-like Coulomb spectroscopy in the absence of zero-bias peaks. *Nature*. **612**, 442–447 (2022).

63. S. M. Albrecht, A. P. Higginbotham, M. Madsen, F. Kuemmeth, T. S. Jespersen, J. Nygård, P. Krogstrup, C. M. Marcus, Exponential protection of zero modes in Majorana islands. *Nature*. **531**, 206–209 (2016).

64. S. A. Khan, C. Lampadaris, A. Cui, L. Stampfer, Y. Liu, S. J. Pauka, M. E. Cachaza, E. M. Fiordaliso, J.-H. Kang, S. Korneychuk, T. Mutas, J. E. Sestoft, F. Krizek, R. Tanta, M. C. Cassidy, T. S. Jespersen, P. Krogstrup, Highly Transparent Gatable Superconducting Shadow Junctions. *ACS Nano*. **14**, 14605–14615 (2020).

65. K. Zuo, V. Mourik, D. B. Szombati, B. Nijholt, D. J. van Woerkom, A. Geresdi, J. Chen, V. P. Ostroukh, A. R. Akhmerov, S. R. Plissard, D. Car, E. P. A. M. Bakkers, D. I. Pikulin, L. P. Kouwenhoven, S. M. Frolov, Supercurrent Interference in Few-Mode Nanowire Josephson Junctions. *Phys. Rev. Lett.* **119**, 187704 (2017).

66. S. R. Mudi, S. M. Frolov, Model for missing Shapiro steps due to bias-dependent resistance (2022), , doi:10.48550/arXiv.2106.00495.